\documentclass[journal]{IEEEtran}

\ifCLASSINFOpdf
\else
   \usepackage[dvips]{graphicx}
\fi
\usepackage{url}
\usepackage{amsmath}
\usepackage{graphicx}
\usepackage{placeins}
\usepackage{float}
\usepackage{bbm}
\usepackage{subfigure}
\usepackage{verbatim}
\usepackage{amssymb}
\usepackage{amsmath}
\usepackage{amsthm}
\usepackage{mwe}
\usepackage{docmute}
\usepackage[ruled, vlined]{algorithm2e}
\usepackage{gensymb}

\usepackage[disable]{todonotes}

\usepackage{graphicx}

\DeclareMathOperator*{\argmin}{argmin}

\newcommand{\taninv}{\tan^{-1}}

\begin{document}

\title{Multi-Target Localization Using Polarization Sensitive Arrays}
\author{William W. Howard, R. Michael Buehrer
\thanks{W.W. Howard and R.M. Buehrer are with Wireless@VT, Bradley Department of ECE, Virginia Tech, Blacksburg, VA, 24061. \\
e-mails:$\{$wwhoward, buehrer$\}$@vt.edu  }}

\maketitle
\pagenumbering{roman}

\begin{abstract}
    In this work we develop clustering techniques for the Bearing Only Target Localization (BOTL) problem. 
    Our scenario has a receiver move along some path, generating bearing estimates on some interval. 
    Multiple emitting targets exist in the environment, and may be physically close with respect to the receiver distance. 
    The first method is iterative and uses only a series of observed bearings to multiple targets to establish clusters. 
    In each iteration, target positions are estimated using a nonlinear least squares solution. 
    The second technique uses additional polarization information, and is shown to be more effective while requiring more information. 
    In addition the second technique is non-iterative and requires far less computation. 
    In this work we presume knowledge of the number of targets. 
    We conclude by providing simulations of our method and show that the proposed approach outperforms previously proposed methods. 
\end{abstract}

\section{Introduction}
Bearing-only target localization (BOTL) is a useful capability for any passive (non-emitting) system which wishes to obtain the relative positioning of emitters in the environment. 
A passive measurements system is one which does not explicitly emit energy into the environment, and instead depends on the environment to provide information on the targets. 
Specifically in this work we concentrate on Direction of Arrival (DoA) estimates, where a receiver uses directional information obtained from a direction-sensitive antenna or array in a waveform emitted by a target to determine the target's direction. 
Other measurements types could include acoustic or optical information. 

Direction of Arrival estimation comes in various forms, from brute-force maximum likelihood techniques to more mathematically rigorous approaches like MUSIC and ESPRIT. 
As DoA estimation has been widely covered in the literature\todo{cite}, we will not discuss it deeply. 

The bearing angle is the angle from true north (or some arbitrary universal direction) to the target from a given measurement location. 
In a noiseless environment, the solution to a BOTL problem can be found as the intersection of all the bearing lines from several measurement locations, with the measurement locations chosen according to some light assumptions. 
If instead we consider a noisy environment, we must turn to maximum likelihood or least squares techniques to find a solution, as the bearing lines will intersect at more than a single point. 

Previous work \cite{Stansfield_DF} investigated the statistical performance of an estimator which determines all of the intersection points of noisy bearing measurements to determine a target location. This method proved inferior to least-squares approximations. 

Closed-form solutions known as the Stansfield estimator and Orthogonal Vector (OV) method were developed for the BOTL problem in \cite{Stansfield_DF, DOGANCAY20051695}. 
Both of these are considered to be linear least squares techniques, which are inherently biased. 
Fortunately they are simple to implement, and provide decent solutions. 
To eliminate the bias inherent to these estimators, techniques such as Instrumental Variable (IV) \cite{DOGANCAY2004487}, Constrained Least Squares (CLS) \cite{HO_asymptotically}, and Total Least Squares (TLS) \cite{GU_power-bearing} were proposed. 
IV and CLS were shown to be asymptotically unbiased, while the TLS solution was shown to be an asymptotically maximum likelihood estimator. 

In addition, nonlinear least squares approximations have been investigated. 
Nonlinear least squares can attempt to more accurately capture the measurement model, where LoBs are determined by an inverse tangent angle. 
In \cite{FOY_taylor-series}, the nonlinear least squares technique is discussed and a Taylor series approximation is obtained. 

BOTL scenarios can largely be categorized into two bins: those using a network of sensors, and those using multiple observations from a moving platform. We'll focus on the latter, specifically an aerial vehicle which moves along some noisy trajectory to observe a collection of four-dimensional DoA measurements: azimuth angle (from true north), depression angle, auxillary polarization angle and polarization phase difference. 

Multi-target algorithms have been investigated before. In target-dense environment, algorithms which use LoB intersections for position estimation will tend to fail due to the formation of \emph{ghost nodes}, where many LoBs intersect in a location with no target. This was addressed in the work by Reed and Buehrer \cite{Reed_cluster}, where each new LoB measurement is clustered based on number of intersections with the LoBs used to form the prior position estimate.
The clustering technique we will discuss here differs from this approach in the metric used to assign new observations to target tracks, the clustering technique, and scenario. 

As opposed to previous clustering methods in the literature, our technique is specifically intended for spatially close targets. 
Techniques that rely on LoB intersections to assign labels to observations are not well suited to this scenario, since the LoBs for both targets will have many intersections, intensifying the ghost node problem. 

\emph{Notation}. We denote sets as cursive capital letters ($\mathcal{A}, \mathcal{E}$) and elements of sets as lower case letters ($a\in\mathcal{A}$). If there is an index for the set, it will appear as a subscript ($e_i\in\mathcal{E}$). $\hat{\text{Hats}}$ indicate an estimate. Negative indicies ($X_{-i}$) indicate the set without the specified index, $\mathcal{X}\backslash X_{i}$

\emph{Contributions}. 
We provide two clustering algorithms for the BOTL problem. 
Both techniques are shown to be effective in various noise conditions. 
In addition we discuss current least squares estimation techniques, and compare several available in the literature. 
We provide a discussion of the tradeoffs between these two methods, as well as numerical simulations to support our conclusions. 

\section{Background}
\subsection{Problem Setup}
Let the environment consist of a single receiver platform along with $N \geq 1$ target emitters. 
While ll of these are have distinct positions in space, the targets may be physically close when compared to the distance to the receiver. 
Each target uses the same electromagnetic frequency, but has a possibly different polarization. 
The receiver is capable of estimating the direction of arrival for all emitters. 

In the $i^{th}$ time step, $1 \leq i \leq T$, let the receiver be at a position $\mathbf{X}_r^i = \left[x_r^i, y_r^i\right]^T$ and let the targets be at positions 

\begin{equation}
    \label{eq:t_pos}
    \mathbf{X}_t = \left[\mathbf{x}_t, \mathbf{y}_t\right]^T = \begin{bmatrix}
    x_1 & x_2 & \dots & x_N\\
    y_1 & y_2 & \dots & y_N
    \end{bmatrix}
\end{equation}
which do not change over time. 

Via some Direction of Arrival estimator, the receiver collects a bearing measurement for each target in the vector $\hat{\boldsymbol{\theta}^i} = \left[\theta_1^i, \theta_2^i, \dots, \theta_N^i\right]$ which is an estimate of the true bearing $\boldsymbol{\theta}^i$. 
The measurements $\hat{\boldsymbol{\theta}^i}$ can be represented as 
\begin{equation}
    \label{eq:measurement}
    \hat{\boldsymbol{\theta}^i} = \boldsymbol{\theta}^i + \mathcal{N}^i = \taninv\left(\frac{\mathbf{y}_t - \mathbf{y}_r^i}{\mathbf{x}_t - \mathbf{y}_r^i}\right) + \mathcal{N}^i
\end{equation}
where $\mathcal{N}^i$ is zero-mean Gaussian noise with variance $\sigma^2$. 

Before continuing we must establish several assumptions relating to the set of all receiver locations $\{\mathbf{X}_r\}$. 
Generally, for any BOTL algorithm to work, we need for the set of measurements to follow two basic assumptions \cite{DOGANCAY2004487}: \begin{itemize}
    \item \emph{\textbf{Assumption 1}} \\
    There is more than one unique measurement location.
    \begin{equation}
        \exists m_i,m_j \in \mathcal{M} \; \text{ such that } \; P_i \neq P_j
    \end{equation}
    
    \item \emph{\textbf{Assumption 2}} \\
    The set $\{P\}$ is not colinear with any target $t_a \in \mathcal{T}$.
\end{itemize}
Otherwise, the receiver will not collect enough information to estimate the target location. 
Given these assumptions we'll assume either a compliant linear receiver trajectory or a circular trajectory centered near the target(s). 

From this series of estimates $\{\hat{\boldsymbol{\theta}}^j | j \leq i\}$, the receiver can use some algorithm $\mathcal{P}(\hat{\boldsymbol{\theta}},\mathbf{X}_r)$ to acquire an estimate of the target positions $\hat{\mathbf{X}_t}$. 

%
%

\subsection{Target Localization}

The BOTL problem has many solutions. We'll discuss a non-linear least squares method using a Levenberg-Marquardt algorithm. 
In a least squares curve fitting problem, we seek the values which minimize an objective function. 
For our case, the objective function is the measurement equation, Eq. (\ref{eq:measurement}). 
Generally speaking, the problem is posed as Eq. (\ref{eq:lsq})
\begin{equation}
    \label{eq:lsq}
    \hat{\beta} = \argmin_{\beta}\left(\sum_{i=1}^m\left[y_i - f(x_i, \beta)\right]^2\right)
\end{equation}
where $\beta$ is the parameter we're trying to find, $\hat{\beta}$ is the local minimum found by the solving algorithm, $y_i$ is the collected data, and $x_i$ is a known variable. 
It should be noted that the Levenberg-Marquardt least-squares solver can only guarantee the finding of a \emph{local} minimum, not global.

In our application, we can substitute the vector of measurements corresponding to a given target for $y_i$. 
Further, since we're estimating target position $\hat{X}_t$, this becomes the space we search over. 
Now we can see that the objective function $f(\cdot)$ is
\begin{equation}
    f(X_r^i, R) = \taninv\left(\frac{\mathbf{y}_t - \mathbf{y}_r^i}{\mathbf{x}_t - \mathbf{y}_r^i}\right)
\end{equation}
where $R$ is the space of target locations.

Now we can rewrite the least squares problem as
\begin{equation}
    \label{eq:lsq2}
    \hat{X}_t^\ell = \argmin_{R}\left(\sum_{j=1}^i\left[\hat{\theta}_\ell^j - f(X_r^j, R)\right]^2\right)
\end{equation}
%

\section{Clustering Algorithms}
In each time step $i$ we have the set of $N$ unlabeled measurements $\hat{\theta}^i$. 
Since each measurement corresponds to a single target, we need some method of labeling the measurements $\hat{\theta}^i$. 
We can form a set $L_\ell$ which contains the indicies of the measurements in each time step which correspond to the $\ell^{th}$ target. In other words, let $L_\ell$ be the set of indicies such that 
\begin{equation}
    \hat{\boldsymbol{\theta}}_\ell = \cup_{j=1}^i\{\hat{\theta}_{L_\ell^j}^j\}
\end{equation}
is the set of all measurements $\hat{\theta}$ that correspond to the $\ell^{th}$ target. 
Now, to form \emph{estimated} labels, we can use a clustering algorithm. 

Algorithms such as $K$-Means exist to form clusters out of data from mixed distributions. 
For instance, if we have a vector $W$ where each element $w_i$ is either sampled from $x_1 \sim \mathcal{N}(\mu_1,\sigma^2)$ or $x_2 \sim \mathcal{N}(\mu_2, \sigma^2)$, then $K$-Means can be used to estimate which elements come from which distribution. 

Since each of our measurements $\hat{\theta}$ are sampled from a different distribution, due to the receiver motion, we are unable to use $K$-means without introducing additional information. 

We'll discuss two clustering methods. 
The first depends only on bearing measurements, and uses an iterative process to assign labels. 
The second method includes polarization measurements, which will not change due to the position of the receiver. 

\paragraph{By Bearing}
Let $\mathcal{P}(\hat{\boldsymbol{\theta}}, \mathbf{X}_r)$ be an unbiased estimator of target position, using a set of measurements $\hat{\boldsymbol{\theta}}$ and a set of receiver positions $\mathbf{X}_r$ as inputs. 
By using a prior estimate of target locations $\mathbf{X}_t^{i-1}$ and the current receiver location $\mathbf{X}_t^i$, we can form the predicted bearing angles $\overline{\theta}^{i}$ from Eq. (\ref{eq:predicted_theta}). 
\begin{equation}
    \label{eq:predicted_theta}
    \overline{\theta}^i = \taninv{\frac{y_{t}^{j-1} - y_r^j}{x_{t}^{j-1} - x_r^j}}
\end{equation}
Then, by using some cost-minimizing algorithm, we can form a one-to-one map from the next measurements $\hat{\theta}^i$ to the target positions $\mathbf{X}_t^{i-1}$. 
To see this, we can first define the angular distance between two angles $\phi_1$ and $\phi_2$ as the Euclidean distance and denote this as $d_{\phi_1, \phi_2}$. 
Then, take the angular distance between each pair of predicted and measured bearing angles and form a matrix. 
\begin{equation}
    \begin{bmatrix}
    d_{\overline{\theta}_1, \hat{\theta}_1} & d_{\overline{\theta}_1, \hat{\theta}_2} & \dots & d_{\overline{\theta}_1, \hat{\theta}_N}\\
    d_{\overline{\theta}_2, \hat{\theta}_1} & \ddots & \dots & \vdots \\
    \vdots & \vdots & \ddots & \vdots \\
    d_{\overline{\theta}_N, \hat{\theta}_1} & \dots & \dots & d_{\overline{\theta}_N, \hat{\theta}_N}
    \end{bmatrix}
\end{equation}
Note that the notation $\overline{\theta}_n$ and $\hat{\theta}_n$ does not imply that $\overline{\theta}_n$ is the predicted measurement for $\hat{\theta}_n$, and the indices merely represent ordering. 
We can then use some cost-minimizing algorithm (e.g. the Hungarian algorithm \cite{Hungarian}, also known as the Munkres assignment algorithm (MAA) \cite{Munkres}) to update the set of estimated labels $L_\ell$ as
\begin{equation}
    L_\ell = [L_\ell, n]
\end{equation}
where $n$ is the measurement index that corresponds to each predicted index $\ell$. Then, the $j^{th}$ element of $L_\ell$ will be the estimated label for $\hat{\theta}^j$. 

Using all previous measurements as well as the new ones, we can form a new estimate of each target's position by using $\mathcal{P}$. 
Algorithm (\ref{algo:one}) formalizes this argument. 

\begin{algorithm}
\SetAlgoLined
\KwResult{$\hat{\mathbf{X}_t^{i}}$} 
Input $[\hat{X}_t]_0^{i-1}, \; [\hat{X}_r]_1^{i}, \; [\hat{\theta}]_1^{i}, \; [\hat{L}]_1^{i}$\\
\vspace{3mm}
    \For{$j=1:i$}{
        \For{$\ell=1:K$}{
            $\overline{\theta}_\ell^j = \taninv{\frac{y_{t,l}^{j-1} - y_r^j}{x_{t,l}^{j-1} - x_r^j}}$
        }
        $\hat{L}^j = \text{MAA}\left(\overline{\theta}^j, \hat{\theta}^j\right)$\\
        \For{$\ell=1:K$}{
            $\hat{X}_{t,\ell} = \mathcal{P}\left(\cup_{k=1}^j\{\hat{\theta}_{L_\ell^k}^k\}, [\mathbf{X}_r]_1^j\right)$
        }
    }
\caption{Iterative Target Localization}
\label{algo:one}
\end{algorithm}

\emph{Initialization. }
We assume that this algorithm is initialized with some target location estimate $\left[\hat{X}_t\right]_0$. 
Alternatively, the algorithm can be self-starting by using K-Means clustering on the first few bearing estimates. 
This works so long as the total receiver displacement during the initialization phase is small (relative to the target distance) such that there is no overlap in bearing angle.

\paragraph{By Polarization}
While this method relies on a less arbitrary array geometry (in other words, a geometry which provides polarization information), it can also provide more accurate results, especially in specific target/receiver configurations. 
Estimation of polarization angles requires use of an array that is polarization sensitive. 
We're specifically assuming the use of a \emph{vector-sensor}, which consists of a dipole triad and a magnetic loop triad and is capable of resolving DoA estimates in a wide aperture with azimuth $\theta \in [0,2\pi]$, elevation $\phi \in [-\pi/2,\pi/2]$, auxiliary polarization angle $\gamma \in [0, \pi/2]$ and polarization phase difference $\eta \in [-\pi, \pi]$. 

For this method to work, the transmitters are assumed to have different polarization angles. 
This assumption is not too restrictive, since it's likely for handheld or vehicle mounted transmitters to be slightly misaligned. 
In addition, due to the fact that different arrays will have different polarization characteristics, we can expect for dissimilar emitters to be identifiable via their polarization angles. 
As we'll show later, we should expect for emitters with more similar polarization angles to require a higher SNR (and therefore a lower measurement variance) to successfully cluster estimates. 

Now, rather than relying on the previous position estimate, we can directly cluster similar polarization angles, since the the expected polarization measurements will not change with different receiver locations. 
\begin{algorithm}
\SetAlgoLined
\KwResult{$\hat{\mathbf{X}_t^{i}}$} 
    Input $[\hat{X}_r]_1^{i}, \; [\hat{\theta}]_1^{i}$\\
        \vspace{3mm}
        $\hat{L}^j = \text{\emph{K}-means}([\gamma, \eta])$\\
        \For{$\ell=1:K$}{
            $\hat{X}_{t,\ell} = \mathcal{P}\left(\cup_{k=1}^j\{\hat{\theta}_{L_\ell^k}^k\}, [\mathbf{X}_r]_1^j\right)$
        }
\caption{\textit{Non}-Iterative Target Localization}
\label{algo:two}
\end{algorithm}
%

\section{Results}
\subsection{Single Target Simulations}
In order to establish expectations for multi-target localization, we'll first discuss the single target case. 
A useful tool in this sort of analysis is the Cramer-Rao Lower Bound (CRLB). 
The CRLB is a statistical lower limit on the variance of any unbiased estimator. 
The CRLB for BOTL problems has been variously derived in the literature. 
\begin{equation}
    \text{CRLB}(\theta) = \sigma \frac{\sqrt{T}}{\sqrt{\left(\sum{\cos^{2}\theta}\sum{\sin^{2}\theta}\right) - \sum{\left(\cos \theta \sin \theta\right)^2}}}
    \label{eq:crlb}
\end{equation}

Without loss of generality, and assuming a linear flight path, we can let the trajectory exist on the positive x-axis and the target be in the first quadrant. 
Were we to put the receiver trajectory on another line, we could transform the coordinates through rotation or translation so that the trajectory moves to the x-axis. 
Since this transformation will not alter the bearing measurements, it will not change the accuracy of an estimator.

To show the accuracy of an estimator, we'll use the root mean squared error (RMSE). 
The RMSE is calculated as
\begin{equation}
    RMSE(\hat{X}, X) = \sqrt{\left(\sum_{j=1}^i (\hat{X}_j - X_j)\right)^2}
\end{equation}

\begin{figure}
    \centering
    \includegraphics[scale=0.6]{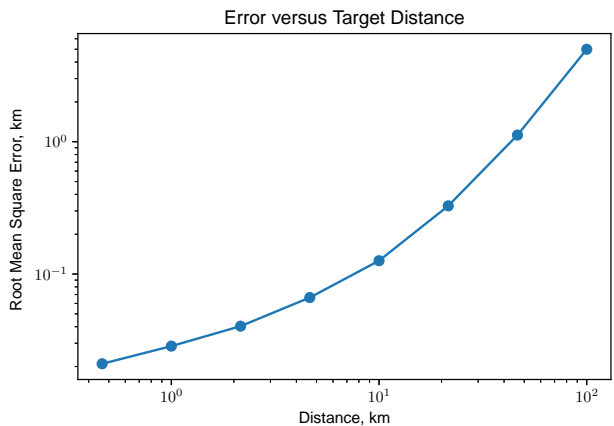}
    \caption{Several trials where the target moves logarithmically further from the receiver plotted on a log-log scale. }
    \label{fig:ysweep}
\end{figure}

We'll first show a series of simulations where a target is progressively further from the x-axis. 
The receiver takes 100 measurements evenly spaced between $x=0$km to $x=30$km. 
The target $x$ location is held at 15km, while the $y$-value is swept as shown. 
In Fig. \ref{fig:ysweep}, we can see that, as we'd expect, the RMSE increases with distance. 

\begin{figure}
    \centering
    \includegraphics[scale=0.6]{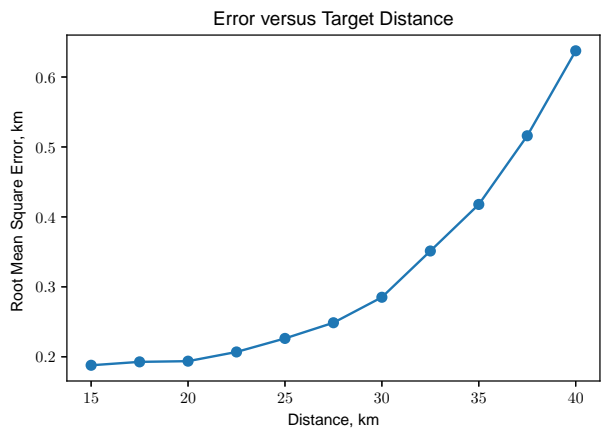}
    \caption{Several trials where the target moves parallel to the receiver trajectory, moving from one end to the other. The receiver makes 100 measurements equally spaced from [0, 0]km to [0, 30] km while the target takes steps from 15km to 40km. RMSE is plotted logarithmically. }
    \label{fig:xsweep}
\end{figure}

Next, we'll look at the performance of the estimator as a target is moved in the $x$ direction while the $y$ coordinate is kept constant. 
The receiver moves in each simulation as described above. 
The target here always has a $y$ value of $15$km, while the $x$ value changes from $15$km to $40$km. 
Fig. \ref{fig:xsweep} shows us that as a target is further from the midpoint of a trajectory, the RMSE degrades. 

\begin{figure}
    \centering
    \includegraphics[scale=0.6]{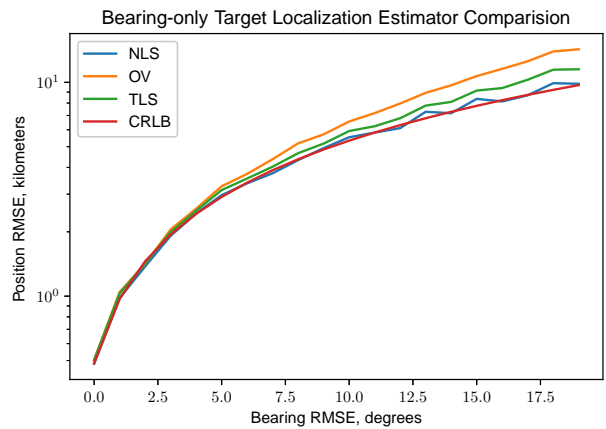}
    \caption{A comparison of several estimators in the literature, with logarithmic position RMSE. Orthogonal Vector (OV) and Total Least Squares (TLS) are prior techniques, while Non-linear Least Squares (NLS) is the technique used for the rest of this work. }
    \label{fig:est_cmp}
\end{figure}

Lastly, we'll see the effect of varying bearing estimator variance. 
Here we'll also look at the OV and TLS estimators, as well as the CRLB. 
Fig. \ref{fig:est_cmp} shows us that the nonlinear least squares estimation outperforms the other estimators. 
In addition, it nearly attains the CRLB. 

\subsection{Multi-Target Simulations}
Now that we have established the performance of our estimator against others in the literature, we'll consider the capabilities and limitations of the clustering techniques we described above. 

\paragraph{Two target orientation}
We'll show that an inherent weakness in the iterative clustering technique is the ability of the method to correctly cluster targets which are oriented perpendicular to the flight path. 
By this, we mean that a line intersecting both targets would be normal to the receiver trajectory. 
This is important in comparison to the parallel target orientation case, where a line intersecting both targets would be parallel to the receiver trajectory. 
Obviously with a circular receiver trajectory, these cases would be symmetrical. 

\begin{figure}
    \centering
    \includegraphics[scale=0.6]{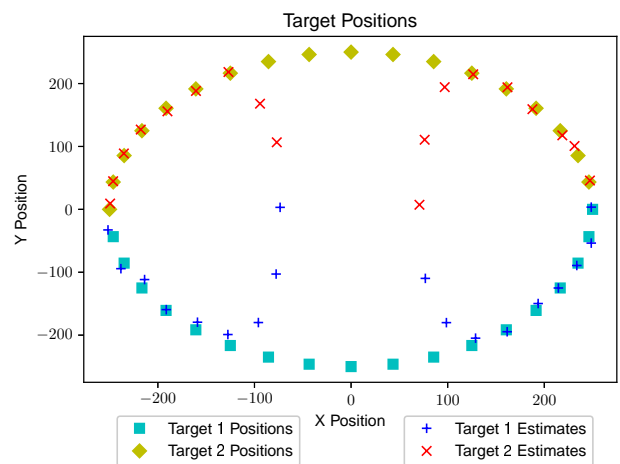}
    \caption{A comparison of estimation accuracy versus target orientation. In each of 18 different simulations, two targets oppose each other while the receiver attempts to cluster observations for localization. The receiver takes a linear path parallel to the x-axis. Algorithm \ref{algo:one} is used for this figure. We can see that when a line through the targets is perpendicular to the flight path, the clustering technique breaks down. }
    \label{fig:bad_cluster}
\end{figure}
\begin{figure}
    \centering
    \includegraphics[scale=0.6]{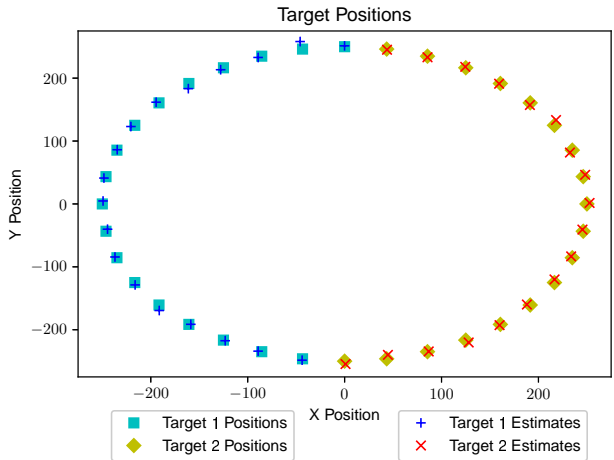}
    \caption{A comparison of estimation accuracy versus target orientation. In each of 18 different simulations, two targets oppose each other while the receiver attempts to cluster observations for localization. The receiver takes a linear path parallel to the x-axis. Algorithm \ref{algo:two} is used for this figure. }
    \label{fig:good_cluster}
\end{figure}

So, we can see in Fig. \ref{fig:bad_cluster} the effect of target orientation on our clustering algorithm. We know that this increased error is due to the clustering algorithm and not the localization algorithm because the separation between targets is much much smaller than the distance from the origin to the receiver, so any rotation effect should not have an impact. 
In each of several simulations, two targets are counter-posed on a circle. Through sequential simulations, the targets remain counter-posed but rotate around the circle. 
As the targets come perpendicular to the receiver, the estimated bearing angles become closer and closer and appear to cross. At this point the future clustering labels become reversed, causing the location estimates to converge.

When we attempt to use the polarization clustering technique, as in Fig. \ref{fig:good_cluster}, we can see that this effect disappears. 
This is because we're no longer using an iterative approach, which relies on the previous location estimate to inform the next cluster estimate. 
Instead, once the data is collected, the polarization information is used to cluster the samples as belonging to one target or the other. 
We can see that this method is superior, as it has fewer requirements on the receiver trajectory.

\begin{figure}
    \centering
    \includegraphics[scale=0.6]{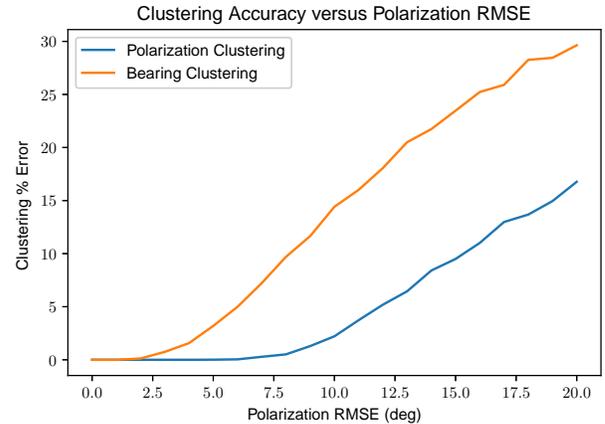}
    \caption{Clustering percent error over a range of bearing estimation error. }
    \label{fig:cluster_error}
\end{figure}

Next, we can look at the effect of varying DoA RMSE on both clustering algorithms. 
For this simulation, we place two targets at $[14500, 15000]$ and $[20500, 15000]$ for an angular separation from the midpoint of the receiver trajectory of 40\degree. 
The polarization parameters for the two targets are also separated by 40\degree. 
Fig. \ref{fig:cluster_error} shows us that at low polarization RMSE (which corresponds to high SNR), we don't see much difference in either clustering technique. 
Note that this assumes a favorable target positioning. 
%

\section{Conclusions}
We have shown that a moving platform collecting bearing measurements can use clustering techniques to localize multiple targets with high accuracy using a nonlinear least squares technique. 
Specifically, we demonstrated two clustering techniques. 
The first requires only bearing angles, and can separate target measurements in a broad variety of situations. However, if the targets are more overlapped from the receiver's perspective, the clustering technique can fail, causing the target localization estimates to become closer together. 

If instead we also have polarization estimates available, we can use that information to cluster received samples. 
We have shown that this method matches the bearing clustering approach's performance in most scenarios, and is superior in the case that the targets are perpendicular to the receiver's trajectory. 

\bibliographystyle{IEEEtran}
\bibliography{bibli}

\begin{thebibliography}{1}
\providecommand{\url}[1]{#1}
\csname url@samestyle\endcsname
\providecommand{\newblock}{\relax}
\providecommand{\bibinfo}[2]{#2}
\providecommand{\BIBentrySTDinterwordspacing}{\spaceskip=0pt\relax}
\providecommand{\BIBentryALTinterwordstretchfactor}{4}
\providecommand{\BIBentryALTinterwordspacing}{\spaceskip=\fontdimen2\font plus
\BIBentryALTinterwordstretchfactor\fontdimen3\font minus
  \fontdimen4\font\relax}
\providecommand{\BIBforeignlanguage}[2]{{%
\expandafter\ifx\csname l@#1\endcsname\relax
\typeout{** WARNING: IEEEtran.bst: No hyphenation pattern has been}%
\typeout{** loaded for the language `#1'. Using the pattern for}%
\typeout{** the default language instead.}%
\else
\language=\csname l@#1\endcsname
\fi
#2}}
\providecommand{\BIBdecl}{\relax}
\BIBdecl

\bibitem{Stansfield_DF}
R.~G. {Stansfield}, ``Statistical theory of {D.F.} fixing,'' \emph{Journal of
  the Institution of Electrical Engineers - Part IIIA: Radiocommunication},
  vol.~94, pp. 762--770, 1947.

\bibitem{DOGANCAY20051695}
\BIBentryALTinterwordspacing
K.~Doğançay, ``Bearings-only target localization using total least squares,''
  \emph{Signal Processing}, vol.~85, no.~9, pp. 1695--1710, 2005. [Online].
  Available:
  \url{https://www.sciencedirect.com/science/article/pii/S0165168405000903}
\BIBentrySTDinterwordspacing

\bibitem{DOGANCAY2004487}
\BIBentryALTinterwordspacing
------, ``Passive emitter localization using weighted instrumental variables,''
  \emph{Signal Processing}, vol.~84, no.~3, pp. 487--497, 2004. [Online].
  Available:
  \url{https://www.sciencedirect.com/science/article/pii/S0165168403003153}
\BIBentrySTDinterwordspacing

\bibitem{HO_asymptotically}
K.~Ho and Y.~Chan, ``An asymptotically unbiased estimator for bearings-only and
  doppler-bearing target motion analysis,'' \emph{IEEE Transactions on Signal
  Processing}, vol.~54, no.~3, pp. 809--822, 2006.

\bibitem{GU_power-bearing}
G.~Gu, ``A novel power-bearing approach and asymptotically optimum estimator
  for target motion analysis,'' \emph{IEEE Transactions on Signal Processing},
  vol.~59, no.~3, pp. 912--922, 2011.

\bibitem{FOY_taylor-series}
W.~H. Foy, ``Position-location solutions by taylor-series estimation,''
  \emph{IEEE Transactions on Aerospace and Electronic Systems}, vol. AES-12,
  no.~2, pp. 187--194, 1976.

\bibitem{Reed_cluster}
J.~D. Reed, C.~R. C.~M. da~Silva, and R.~M. Buehrer, ``Multiple-source
  localization using line-of-bearing measurements: Approaches to the data
  association problem,'' in \emph{MILCOM 2008 - 2008 IEEE Military
  Communications Conference}, 2008, pp. 1--7.

\bibitem{Hungarian}
H.~W. Kuhn, ``The hungarian method for the assignment problem,'' \emph{Naval
  Research Logistics Quarterly}, pp. 83--97, 1955.

\bibitem{Munkres}
J.~Munkres, ``Algorithms for the assignment and transportation problems,''
  \emph{Journal of the Society for Industrial and Applied Mathematics}, pp.
  32--38, 1957.

\end{thebibliography}
\end{document}